\documentclass[%
 reprint,
 amsmath,amssymb,
 aps,
]{revtex4-2}

\usepackage{tikz} % Package for drawing
\usepackage{amsmath,amsfonts,amsthm, mathtools}
\usepackage{float}
\usepackage{dsfont}
\usepackage{csquotes}
\usepackage{amssymb}
\usepackage{verbatim}
\usepackage{bm}
\usepackage{hyperref}
\definecolor{darkred}  {rgb}{0.5,0,0}
\definecolor{darkblue} {rgb}{0,0,0.5}
\definecolor{darkgreen}{rgb}{0,0.5,0}
\hypersetup{
  colorlinks = true,
  urlcolor  = blue,         % color of external links
  linkcolor = darkblue,     % color of internal links
  citecolor = darkgreen,    % color of links to bibliography
  filecolor = darkred       % color of file links
}

\theoremstyle{definition}

\newtheorem*{remark*}{Remark}

\usepackage{multirow}

\definecolor{cool_green}{rgb}{0.0, 0.5, 0.0}

\usepackage{blkarray}
\usepackage{graphicx}% Include figure files
\usepackage{dcolumn}% Align table columns on decimal point
\usepackage{bm}% bold math
\usepackage{xcolor}

\usepackage{xcolor}

\newif\ifshowannotation
\showannotationtrue 
\newif\ifshowupdate
\showupdatetrue      
\begin{document}

\preprint{APS/123-QED}

\title{Sleeping Beauty in One or Many Worlds: A Defense of the Halfer Position}

\author{Jiaxuan Zhang}
\email{jiaxuan.zhang@physics.ox.ac.uk}

\affiliation{
Department of Physics, University of Oxford, Oxford, United Kingdom
}

\date{\today}% It is always \today, today,
             %  but any date may be explicitly specified

\begin{abstract}

The Sleeping Beauty Problem (SBP) is a long-standing puzzle in classical probability theory and has been used to challenge the Many-Worlds Interpretation (MWI) of quantum mechanics, since both involve objective determinacy combined with subjective uncertainty about certain events. A common concern is that MWI yields a different answer to the quantum version of SBP than the widely supported Thirder position in the classical case. We argue that this concern is unwarranted. We show that in both the quantum and classical versions of SBP, the correct credence is given by the Halfer position.

In the quantum (MWI) SBP, we show that if no unjustified renormalization is introduced, the correct credence is $1/2$. We then extend this result to the classical SBP by refuting four major arguments for $1/3$. First, we reject the Proportion Argument by distinguishing event weight from probability. Second, we rebut Elga’s Variant Argument by extending an earlier critique that identifies the implicit introduction of additional information; we further clarify this point by constructing a new variant and explaining why the Principle of Indifference is inapplicable, drawing an analogy with a mistake by d’Alembert in the history of probability theory. Third, we identify a flaw in the Technicolor Beauty Variant Argument, which arises from treating overlapping events as disjoint. Finally, we argue that causal decision theory is inappropriate for SBP, rendering the Thirders vulnerable to a Dutch Book.

Our results support the consistency of MWI under the challenge posed by SBP and suggest that the dominant position on SBP needs careful reconsideration.

\end{abstract}

%\keywords{Suggested keywords}%Use showkeys class option if keyword
                              %display desired
\maketitle

%\tableofcontents

% The \nocite command causes all entries in a bibliography to be printed out
% whether or not they are actually referenced in the text. This is appropriate
% for the sample file to show the different styles of references, but authors
% most likely will not want to use it.

\section{Introduction}

The measurement problem is one of the most notorious and critical unresolved problems in quantum mechanics. A central aspect of this problem concerns the emergence of probability. Among quantum theory and all other physical theories, the quantum measurement process is the only process that involves objective probabilities. By objective probability, we distinguish the probabilities in quantum mechanics --- which appears to be intrinsic and to arise from nowhere (at least in the orthodox Copenhagen Interpretation) --- from the subjective probabilities in classical physics, which arise from our ignorance of a system’s detailed microscopic dynamics \cite{Norsen2017}.

The problem of the origin of probability becomes even more urgent in interpretations such as the Many-Worlds Interpretation (MWI), which deny the existence of real random processes \cite{KENT_1990,deutsch1999,LEWIS20071,Zurek_2018}. Roughly speaking, MWI is an interpretation of quantum mechanics in which there is nothing beyond unitary evolution, and a quantum measurement is merely an entanglement between the system and the measuring apparatus. An observer sees only one measurement outcome (or eigenstate) because only one copy of her is split into the same ``world" as that outcome. This also implies that, in MWI, every outcome occurs in reality \cite{saunders2010}. Hence, when we say that a particular outcome $a_i$ has probability $P(a_i)$ as given by the Born rule, this does not mean that $a_i$ will be observed with probability $P(a_i)$. Somewhat surprisingly, every outcome $a_i$ will be observed with probability 1, albeit in different worlds. What $P(a_i)$ really represents is the credence of the observer that she will become the particular copy of herself observes $a_i$.

In this work, we denote the objective probability of an event occurring by $\tilde{P}$, and the credence that a given observer sees it --- or equivalently, that the observer is split into the same world as this event --- by $P$. Supporters of MWI usually prefer to get rid of the concept of observer. Indeed, the observer does not contribute to any special dynamics in MWI \cite{saunders2010}. However, when it comes to probability, the notion of an observer remains useful and even essential.

A very interesting analogy has been noted between probability in MWI and a well-known paradox in classical probability theory: the Sleeping Beauty Problem (SBP) \cite{Lewis.P2007, Hallakoun2013}, which we will introduce in detail in Section \ref{section_intro_SBP}. In both cases, there exists an event $A$ for which $\tilde{P}(A) = 1$ but $P(A) < 1$. It is therefore natural to expect that a quantum version of SBP, interpreted by MWI, should yield the same answer as the classical SBP. This expectation has led some authors to express concern about MWI, arguing that the quantum (MWI) SBP produces an answer that differs from the dominant answer of the classical SBP \cite{Lewis.P2007}. As a reply to this worry, some authors have attempted to derive the dominant answer from the quantum SBP \cite{Hallakoun2013}. Others have attempted to show that the quantum (MWI) SBP and the classical SBP are not genuinely analogous, and therefore should yield different answers \cite{Peterson2011,Wilson2014}. Darren Bradley has proposed a proof that the two situations should both yield the same answer, which is not the dominant one but rather the second most supported answer \cite{Bradley2011-BRACIB}. In this paper, we support Bradley’s position by providing a new and more convincing proof.

The structure of this paper is as follows: In Section \ref{section_intro_SBP}, we introduce the setup of the Sleeping Beauty Problem and review the main competing answers and their respective arguments. In Section \ref{section_refute_1/3}, we refute the Thirders’ arguments, beginning with the quantum SBP and followed by four important proofs for the classical case. We will show how $P(H) = 1/2$ emerges as the only consistent and acceptable answer across all these analyses.

\section{Sleeping Beauty Problem and Positions}
\label{section_intro_SBP}

\subsection{Sleeping Beauty Problem (SBP)}

Sleeping Beauty agrees to participate in the following experiment. She will be put to sleep on Sunday. Then, the experimenter will toss a fair coin. If the coin lands Heads, she will be woken up on Monday; if it lands Tails, she will be woken up on both Monday and Tuesday. Each time she is awakened, the experimenter will give her a pill and put her back to sleep in order to make her forget the previous awakening. When she wakes up, she is asked what her credence that the coin landed Heads ($P(H) = P(H|Wake)$) is \cite{elga2000}.

In quantum SBP, the coin is replaced by a quantum coin, which can be in a superposition of Heads and Tails states \cite{Hallakoun2013}.

The two major answers proposed for the problem are $P(H)=1/3$ and $P(H)=1/2$. In this work, we focus mainly on the debate between these two views and argue in favor of the latter. We will summarize the arguments that we are going to discuss or refute in this section, while the detailed analysis is presented in Section \ref{section_refute_1/3}. Several less supported positions are also briefly introduced and discussed at the end of this section.

\subsection{Thirder Position}

The dominant position claims that $P(H)=1/3$. This view is called the Thirder Position, and those who endorse it are called Thirders.

The most important representative of the Thirders is Adam Elga, who is also one of the first authors to introduce the SBP in the literature \cite{PICCIONE19973,elga2000}. Elga proposes two proofs for $P(H)=1/3$ \cite{elga2000}:

(1) Proportion Argument --- If the experiment is repeated many (for example, infinite) times, and we count all of Beauty’s wakings, the frequency of wakings following a coin landing Heads divided by the total number of wakings will be $1/3$.

(2) Elga's Variant Argument --- Elga argues that the credences satisfy $P(T \cap Mo) = P(T \cap Tu)$ and $P(H \cap Mo) = P(T \cap Mo)$, where $H,T$ denote the events {``The coin lands Heads", ``The coin lands Tails"}, and $Mo, Tu,$ and $Su$ denote {``It is Monday now", ``It is Tuesday now", ``It is Sunday now"}, respectively. The first equality is trivial, and there is no debates on it, while the second requires further explanation and is more controversial \cite{Lewis2001,Burock2023}.

We will also refute a proof based on another variant called Technicolor Beauty, introduced by Michael Titelbaum \cite{Titelbaum2008-TITTRO}. This variant involves introducing an independent ancillary experiment, and many people believe that $P(H)$ remains the same as in the original SBP and regard this proof as very convincing \cite{Briggs2010-BRIPAV-5,Wilson2014,Conitzer_2015,Pittard2015}.

Another method usually used to test whether a credence is reasonable is to construct a Dutch Book against it. There are Dutch Books against both $P(H)=1/3$ and $P(H)=1/2$ \cite{halpern2004sleepingbeautyreconsideredconditioning,Hitchcock2004-HITBAT,Briggs2010-BRIPAV-5}. We will discuss them in detail in Section \ref{subsection_Dutch_Book}.

Finally, for the quantum version of the SBP, Peter Lewis was the first to raise the issue and to prove $P(H)=1/2$ by constructing an equivalence between the situation in which Beauty wakes up on both days if the coin lands Tails and another situation in which the experimenter tosses a second quantum coin and wakes her up on only one of the days. Despite this argument, Peter Lewis prefers the Thirder position to the Halfer position. The aim of his proof is to demonstrate an incoherence between the results obtained from the classical SBP ($P(H)=1/3$) and the quantum (MWI) SBP ($P(H)=1/2$) \cite{Lewis.P2007}. Other authors share this concern and attempt to resolve the problem: Na'ama Hallakoun, Berry Groisman, and Lev Vaidman argue that Lewis' variant is not equivalent to the original setup and therefore changes the relevant credence. They then derive $P(H)=1/3$ using the relationship between the amplitudes of quantum eigenstates and probability \cite{Hallakoun2013}. In Section \ref{subsection_Quantum_SBP}, we will show that there is no need to worry about the MWI from this perspective for a different reason: we will prove that, in both the quantum and classical cases, $P(H)=1/2$ is the only correct answer. 

We will go through and refute all the proofs mentioned above for the Thirder position in Section \ref{section_refute_1/3}, starting with the quantum version and followed by the classical ones.

\subsection{Halfer Position}

The next most supported position claims that $P(H)=1/2$. It is called the Halfer Position, and people who endorse it are called Halfers.

The most important representative of the Halfers is David Lewis. His reasoning is that no new updating information is gained when Beauty wakes up, since she will be awakened regardless of the result of the coin toss. Therefore, she should not change her credence $P(H)$ \cite{Lewis2001}. We can make this point clearer using Bayes' Theorem:

\begin{align*}
&~P(H_{Su}|Wake) = \frac{P(Wake|H_{Su}) \cdot P(H_{Su})}{P(Wake)} = \frac{1 \cdot P(H_{Su})}{1} \\
&~~~~~~~~~~~~~~~~~~~~ = P(H_{Su}) \\  
&~~~~~~~~~~~~~~~~~~~~ = \frac{1}{2} \\  
\end{align*}

Interestingly, Elga also endorses the use of conditional probability. In his paper, he argues that $P(H)=1/3$ is reasonable by showing that it is a weighted average of $P(H|Mo)$ and $P(H|Tu)$ \cite{elga2000}:

$$P(H)=P(H|Mo) \cdot P(Mo) + P(H|Tu) \cdot P(Tu) = \frac{1}{2} \cdot \frac{2}{3} + 0 = \frac{1}{3} $$

This weighted-average explanation is also applicable to the Halfer position. We list the calculation here because we will use these probabilities repeatedly later:

$$P(H)= \frac{2}{3} \cdot \frac{3}{4} + 0 = \frac{1}{2} $$

The key problem we want to point out here, however, is that if conditional probability is valid, then for Thirders like Elga it implies that

$$\frac{P(Wake|H_{Su})}{P(Wake)} = \frac{2}{3}$$

This seems to imply that if the coin lands Head, Beauty is not guaranteed to be awakened, which is a very strange result and appears inconsistent with the SBP experimental setup.

We will not further explain proofs for $P(H)=1/2$ in this work. Rather, in Section \ref{section_refute_1/3}, we will show that for all proofs proposed for $P(H)=1/3$, the correct answer should instead be $1/2$.

\subsection{Other Viewpoints}

We shall also briefly review two other less-supported but still interesting positions here: the Double Halfer Position and the Ambiguous Question Position.

The Double Halfers believe that $P(H)=1/2$ and $P(H|Mo)=1/2$. This implies that the conditional probability updating is not applicable to SBP. Mikaël Cozic argues that an alternative rule, the imaging rule, should be applied to SBP \cite{COZIC2011137}. We doubt this viewpoint. Because in probability theory, imaging rules are only preferable to the conditional probability updating when there exist some real changes in reality. Examples include the change of time on a clock or the removal of some balls from a box. Cozic claims that the change of date is this kind of real change \cite{COZIC2011137}. However, in SBP, we are asking about the credence for the result of the coin toss rather than the credence for the date. The coin toss already happens on Sunday and never changes afterward. Therefore, we think the imaging rule is not proper here, and the conditional probability updating is still the best choice.

Additionally, as a supporter of the Ambiguous Questions Position, Berry Groisman gives a very nice example showing how 1/2 and 1/3 can both be correct answers to different questions. Suppose the experimenter puts a green ball into a box when the coin lands Heads, and two red balls into the same box when the coin lands Tails. After repeating the experiment many times, if we ask “what is the credence for putting a green ball in the box,” the correct answer should be 1/2; and if we ask “what is the credence for picking out a green ball from the box,” the correct answer should be 1/3 \cite{groisman2008endsleepingbeautysnightmare}. 

This is a clear and insightful illustration. However, we do not agree with Groisman’s argument that the original SBP question is closer to “picking out a green ball.” We think that the “picking out” question is based on a new probability distribution in a second experiment, which is generated by the result of the first SBP experiment. Compared with the analysis of the Double Halfer position above, we can see that the real situation is changed here. For example, if we toss the coin 100 times in a very ideal case, in the first SBP experiment the total number of trials is 100; but when we pick out the green balls, half of the trials will be duplicated and the total number of trials becomes 150. Instead, we think the original SBP question is closer to the “putting in a green ball” question.

\section{Refuting the Thirder Position}
\label{section_refute_1/3}

\subsection{Quantum Sleeping Beauty}
\label{subsection_Quantum_SBP}

We have already mentioned the analogy between MWI and SBP concerning cases in which $\tilde{P}(A)=1$ but $P(A)<1$ for some event $A$. Now we take a closer look at the proof for $P(H)=1/2$ raised by Peter Lewis \cite{Lewis.P2007}:

Consider two situations. The first is when the coin lands Tails in the original SBP. In this case, Beauty is woken up on both Monday and Tuesday. We call this the SBP Tail experiment. The second situation still concerns when the coin lands Tails in SBP, but then, a second quantum coin is tossed to decide which day Beauty is woken up: if the second coin is Heads, the experimenter wakes Beauty up on Monday; if the second coin is Tails, the experimenter wakes her up on Tuesday. We call this the Second Q-Toss experiment.

Peter Lewis claims that these two situations can be treated as equivalent, if the quantum coin toss is interpreted by MWI. Then $P(H)=1/2$ follows trivially from the two coin tosses \cite{Lewis.P2007}. We agree with this view. Notice, of course, that the two situations cannot be totally equivalent. For example, there are two worlds in the Second Q-Toss experiment, but only one world in the SBP Tail experiment. However, if we only care about Beauty’s credence, then in both cases we have $\tilde{P}(Mo)=1$ and $P(Mo)=1/2$. Moreover, unlike the problem with Elga’s Variant (see Section \ref{subsection_Elga's_Variant}), Beauty does not obtain any crucial new information from the Second Q-Toss experiment that is lacking in the SBP Tail experiment: whether she wakes up on Monday or Tuesday, she knows that the two coins have both been tossed.

However, some people disagree with this view \cite{Papineau2009-PAPATA,Hallakoun2013}. We just briefly rephrase the argument of authors of \cite{Hallakoun2013} here: They claim that in the SBP Tail experiment, Beauty’s probability measures for events $H$ and $T$ are both 1, and are then normalized to 1/2; whereas in the Second Q-Toss experiment, Beauty’s measures for both events are exactly 1/2 --- This value comes from taking the modulus square of the amplitude in front of an event in a quantum superposition, that is, the result given by the Born rule. 

We think this argument is wrong, because the latter measure should also be 1 and is normalized at the stage of writing down the quantum state. This normalization should not be ignored, nor can it be renormalized again. This double normalization is how the authors of \cite{Hallakoun2013} obtain $P(H)=1/3$.

In \cite{Hallakoun2013}, they write down the superposition states for Monday and Tuesday as follows:

\begin{align*}
|\psi(Mo)\rangle &= \frac{1}{\sqrt{2}}|T \cap Mo: \text{Wake}\rangle + \frac{1}{\sqrt{2}}|H \cap Mo: \text{Wake}\rangle \\
|\psi(Tu)\rangle &= \frac{1}{\sqrt{2}}|T \cap Tu: \text{Wake}\rangle + \frac{1}{\sqrt{2}}|H \cap Tu: \text{Sleep}\rangle
\end{align*}

They then claim that since the amplitudes of all four events are the same, these events have equal probability measures. When Beauty wakes up, she can delete the last event, $H \cap Tu$, and assign equal probability 1/3 to the remaining three events \cite{Hallakoun2013}.

There is an obvious renormalization in this proof. And more importantly, in MWI, the modulus square of the amplitude in front of an eigenstate does not represent the probability of a certain event, but the probability of a certain world. There is only one split of the world resulting from the single quantum coin toss in the original SBP. So the amplitudes actually means that the Heads world and the Tails world both have probability 1/2. Therefore, we surely get $P(H)=1/2$. We even do not need Peter Lewis’ proof to obtain this result: $P(\text{Heads World})=1/2$ is the probability calculated directly from the Born rule.

\subsection{Proportion Argument}
\label{subsection_Proportion_Argument}

The Proportion Argument seems very convincing. However, it contains a hidden error: it takes the weight of a possible event as the probability of it. Consider another game in which we win $1$ if the coin lands Heads, and $2$ if it lands Tails (Monetary units are omitted for simplification in this paper). After 100 games, in a very ideal case, we obtain $300$ in total. We also know that one third of this total reward comes from Heads. It is clear that we should not say that our credence is $P(H)=1/3$, because Heads and Tails do not yield equal rewards.

In SBP, the situation is similar. Although the proportion of times that Heads occurs relative to the number of wakings is 1/3, we should not conclude that Heads occurs 1/3 of the time. We should calculate $P(H)$ using the number of experiments as the denominator, not the number of wakings.

\subsection{Elga's Variant}
\label{subsection_Elga's_Variant}

Elga completes his proof with two equalities: $P(T \cap Mo) = P(T \cap Tu)$ and $P(H \cap Mo) = P(T \cap Mo)$. Since these are the only three possible events, he then concludes that $P(H \cap Mo)=P(T \cap Mo) = P(T \cap Tu) = 1/3$. The first equality, $P(T \cap Mo) = P(T \cap Tu)$, comes from the Principle of Indifference and is generally not controversial. However, the second equality, $P(H \cap Mo) = P(T \cap Mo)$, has been more widely questioned in the literature \cite{Lewis2001,Burock2023}. We first go through Elga's proof below.

Elga also bases his proof on an “equivalent” setup to the original SBP, which he calls “Method (2)”. In Method (2), instead of tossing the coin on Sunday, the experimenter always wakes Beauty up on Monday and then tosses the coin on Monday evening. The experimenter then wakes her up on Tuesday if the coin lands Tails, or keeps her asleep if it lands Heads.

It is obvious that Method (2) generates the same experimental results and probability distribution as the original SBP. Elga then claims that Beauty’s credence is the same in the two setups. From Method (2), if Beauty is told that it is Monday, her conditional credence for Heads and Tails should be the same, since the coin has not yet been tossed:

$$P(H \cap Mo \mid Mo) = P(T \cap Mo \mid Mo)$$

By Bayes' Theorem, we have

\begin{align*}
P(T \cap Mo \mid Mo) &= \frac{P(Mo \mid T \cap Mo) P(T \cap Mo)}{P(Mo)} \\
P(H \cap Mo \mid Mo) &= \frac{P(Mo \mid H \cap Mo) P(H \cap Mo)}{P(Mo)} \\
\end{align*}

So

$$
P(H \cap Mo) = P(T \cap Mo)
$$

This may look convincing at first glance. However, Marc Burock has pointed out that although the probability distributions in the two setups are identical, the information obtained by Beauty is actually different. In Method (2), when Beauty is told that it is Monday, she knows that the coin has not been tossed, whereas in the original SBP, she always believes that the coin has already been tossed. She actually obtains $P(H \cap Mo \mid Mo) = P(T \cap Mo \mid Mo) = 1/2$ from the information that the coin has not been tossed, which is lacking in the original setup \cite{Burock2023}.

We can further question Elga’s proof by slightly adjusting Method (2) to Method (2'): Instead of tossing the coin on Monday evening, the experimenter still tosses the coin on Sunday evening, but only to decide whether Beauty will be woken up on Tuesday. It is obvious that Method (2') also has the same probability distribution as Method (2) and the original SBP, and this setup is closer to the original SBP than Method (2).

Starting from Method (2'), we can clearly see that we can no longer obtain $P(H \cap Mo \mid Mo) = P(T \cap Mo \mid Mo) = 1/2$:

$$
P(Mo) = P(H \cap Mo)+P(T \cap Mo)=\frac{1}{2} + \frac{1}{4} = \frac{3}{4}
$$

$$
P(H \cap Mo \mid Mo) = \frac{P(Mo \mid H \cap Mo)P(H \cap Mo)}{P(Mo)} 
$$

$$= \frac{1 \cdot \frac{1}{2}}{\frac{3}{4}} = \frac{2}{3}$$

$$
P(T \cap Mo \mid Mo) = \frac{P(Mo \mid T \cap Mo)P(T \cap Mo)}{P(Mo)} 
$$

$$
= \frac{1 \cdot \frac{1}{4}}{\frac{3}{4}} = \frac{1}{3}
$$

We also want to point out an interesting comparison between the SBP and another debate in the history of probability theory: During the early development of probability theory, d’Alembert once proposed the idea that when tossing two fair coins, the probability of getting at least one Heads is $2/3$, rather than $3/4$. This is obviously nonsense from a modern perspective. However, d’Alembert’s argument is in some ways similar to the arguments advanced by Thirders today. He argued that once Heads is obtained on the first toss, the game ends. Only if the coin lands Tails does one proceed to the second toss, which can then result in Heads or Tails. So there are three different possible events in total, and due to a lack of further knowledge ---namely, the Principle of Indifference (which had not yet been proposed) --- each event is assigned a probability of $1/3$ \cite{d_Alembert1754}.

We can see a misuse of the Principle of Indifference in d’Alembert’s argument, Method (2'), and the original setup of the SBP. In d’Alembert’s coin tosses, we know that the event “the first toss is Heads” has probability $1/2$, while the other two events, $TH$ and $TT$, each have probability $1/4$. The Principle of Indifference is not applicable here. Returning to the SBP, Beauty can believe that $P(T \cap Mo)=P(T \cap Tu)$. However, it is not reasonable for her to believe that $P(H \cap Mo)=P(T \cap Mo)$, because she knows from the experimental setup and the previous equality that $P(H \cap Mo)=1/2$ and $P(T \cap Mo)=1/4$. She therefore has sufficient information to refuse the use of the Principle of Indifference here. Laplace himself made this point clear in his book when introducing the Principle of Indifference: if the available information already provides non-equal probabilities, the Principle of Indifference cannot be applied trivially \cite{Laplace1814} 

Thus, we cannot conclude that $P(H\cap Mo \mid Mo) = P(T\cap Mo \mid Mo) = 1/2$ in Method (2') or in the original SBP. Method (2) yields this result only by introducing crucial additional information. This refutes Elga’s proof.

Additionally, note that we are not claiming that waking Beauty up twice is equivalent to a second (classical) coin toss. In the SBP, $P(H)=1/3$ and $P(Mo)=2/3$ may seem reasonable when considering frequencies. However, as we have already shown in Section \ref{subsection_Proportion_Argument}, this proportion is misleading and arises from non-equal weights of different possible events.

\subsection{Technicolor Beauty}
\label{subsection_Technicolor_Beauty}

We will refute another proof for the Thirder position which also includes the use of a variant, Technicolor Beauty. In this variant, Michael Titelbaum introduces a friend of Beauty who will throw a die on Sunday evening. If he gets an odd number, he will show Beauty a red paper on Monday and a blue paper on Tuesday; if he gets an even number, he will show Beauty a blue paper on Monday and a red paper on Tuesday. Notice that if Beauty does not wake up on Tuesday, the friend cannot show her anything. The original SBP experiment is conducted normally \cite{Titelbaum2008-TITTRO}.

According to Titelbaum, the die roll is independent of the coin toss, so it will not change Beauty’s credence $P(H)$. He then uses this variant to argue that $P(H)=1/3$: there are only four possible events, {$HO$, $HE$, $TO$, $TE$} (where $O$ and $E$ indicate an odd number and an even number on the die, respectively), each with probability $1/4$.

\begin{table}[h]
    \centering
    \begin{tabular}{|c|c|c|c|}
    \hline
 & P &  Mo & Tu  \\
\hline
HO & 1/4 & Red &  \\
\hline
HE & 1/4 & Blue &  \\
\hline
TO & 1/4 & Red & Blue \\
\hline
TE & 1/4 & Blue & Red \\
\hline
    \end{tabular}
    \caption{All possible events and their corresponding probabilities (P) in the Technicolor Beauty experiment.}
    \label{}
\end{table}

If Beauty wakes up and sees the red paper, she can rule out the event $HE$. Thus,

$$P(H|Red)=\frac{\frac{1}{4}}{\frac{3}{4}} = \frac{1}{3}$$

$P(H|Blue)= 1/3$ for the same reason. Therefore, no matter what Beauty sees, she will conclude that $P(H)=1/3$ \cite{Titelbaum2008-TITTRO}.

Titelbaum’s proof is considered convincing by many people \cite{Briggs2010-BRIPAV-5,Wilson2014,Conitzer_2015,Pittard2015}. However, there is a hidden mistake in this argument. It is true that $P(H|Red)=P(H|Blue)= 1/3$, but $P(Red) + P(Blue) \ne 1$. The events $Red$ and $Blue$ are not disjoint. This is clear from the setup: Beauty sees both the red paper and the blue paper if the coin lands Tails. In fact, we have

$$P(Red)=P(Blue)= P(H)\cdot \frac{1}{2} + P(T)= \frac{1}{4}+\frac{1}{2}=\frac{3}{4}$$

and

$$
P(H|Red) = \frac{1}{3} = = \frac{P(\text{Red} \mid H) \cdot P(H)}{P(\text{Red})} = \frac{\frac{1}{2} \cdot P(H)}{\frac{3}{4}}
$$

$$
P(H)=\frac{1}{2}
$$

Thus, we still obtain $P(H)=1/2$ in the Technicolor Beauty experiment.

Another way to prove the same result is to enforce $P(Red)=P(Blue)= 1/2$. This requires $P(Red|T)=P(Blue|T)= 1/2 \ne 1$. This idea is similar to the proof proposed by Darren Bradley. Although $P(Red|T)=1/2$ appears counterintuitive, Bradley attempts to justify it using an argument based on selection effects \cite{Bradley2011-BRACIB}. However, Alastair Wilson claims that Bradley’s method is valid only for the quantum (MWI) SBP and not for the classical SBP \cite{Wilson2014}. Using the new argument that does not rely on any complex explanation of selection procedures, we have shown that $P(H)=1/2$ should be the only reasonable answer for both the Technicolor Beauty experiment and the original SBP.

\subsection{Dutch Book}
\label{subsection_Dutch_Book}

The Dutch Book Theorem is usually used to test whether a credence is rational. In short, the theorem states that a game, which we refer to as a Dutch Book, can be constructed against a player in which she is guaranteed to suffer a loss if her credence is not rational. In most cases, rationality means that the credence is equal to the objective probability \cite{Kyburg1978,Briggs2010-BRIPAV-5}. We also want to point out that the guaranteed loss means that it is independent of any possible events. For example, if the player is asked to toss a coin, whether it lands Heads or Tails, she will lose money. However, she can still avoid sure loss by refusing to play the game (which may contradicts her credence, but she is still able to do that). This clarification will be important for our discussion below.

Different Dutch Books have been constructed against both the Thirder and the Halfer positions, but it seems to many that the Dutch Book attacking $P(H)=1/2$ is more convincing. Christopher Hitchcock has refuted the Dutch Book attacking $P(H)=1/3$ supported by Joseph Halpern by showing there is deception involved \cite{Hitchcock2004-HITBAT,halpern2004sleepingbeautyreconsideredconditioning}. He then presents another Dutch Book attacking $P(H)=1/2$ \cite{Hitchcock2004-HITBAT}. We demonstrate the formulation of Hitchcock's Dutch Book as presented by Rachael Briggs \cite{Briggs2010-BRIPAV-5} in Table \ref{Table_Hitchcock's_Dutch_Book}. Beauty will be asked if she accept the games each time she wakes up.

\begin{table}[h]
    \centering
    \begin{tabular}{|c|c|c|c|}
    \hline
Game & & Heads &  Tails  \\
\hline
1 & Sunday & -15+2$\epsilon$ & 15+$\epsilon$   \\
\hline
2 & Monday/Tuesday & 10+$\epsilon$ & -10+$\epsilon$   \\
\hline
    \end{tabular}
    \caption{Briggs’ expression of Hitchcock’s Dutch Book \cite{Briggs2010-BRIPAV-5}. $\epsilon>0$ is a small amount that gives Beauty a preference when deciding whether to accept the games. Game 1 is necessary because a Dutch Book requires a sure loss regardless of whether the coin lands Heads or Tails.}
    \label{Table_Hitchcock's_Dutch_Book}
\end{table}

According to Hitchcock, both games are coherent with a credence of $1/2$. However, if Beauty is a Halfer and accepts both games, she will incur a guaranteed loss \cite{Hitchcock2004-HITBAT}. If the coin lands Heads, she gains

$$-15+ 2 \epsilon +10 + \epsilon=-5 + 3 \epsilon$$

and if the coin lands Tails, she gains

$$15 + \epsilon -20 + 2\epsilon =-5 + 3 \epsilon$$

Briggs comments that Beauty will be vulnerable to this Dutch Book only if she is both a Halfer and a causal decision theorist, extending an earlier idea by Frank Arntzenius \cite{Arntzenius2002}. In the scenarios we are discussing, being a causal decision theorist, or acting according to causal decision theory (CDT), means that Beauty believes that the other herself waking up on another day may make any possible decision. In contrast, being an evidential decision theorist, or acting according to evidential decision theory (EDT), means that she believes the other herself will make the same decision as the one she is making now. Briggs also comments that this assumption is plausible, since the same factors that influence her decision on Monday will also influence her decision on Tuesday \cite{Briggs2010-BRIPAV-5}.

According to Arntzenius, the key point is that Beauty having a credence of $P(H)=1/2$ is not equivalent to being willing to accept a game at betting odds of 1:1. Although this may appear counterintuitive, it is reasonable in our scenario. If Beauty is an evidential decision theorist, she knows that her choice is not only responsible for the game today, but also for the game on the other day. This means that Game 2 should be updated to a new version, shown in Table \ref{Table_Hitchcock's_Dutch_Book_evidential}. In this case, it is obvious that she should not accept the game if she is a Halfer \cite{Arntzenius2002,Briggs2010-BRIPAV-5}.

\begin{table}[h]
    \centering
    \begin{tabular}{|c|c|c|c|}
    \hline
& Heads &  Tails  \\
\hline
Monday/Tuesday & $10+\epsilon$ & $-20+2\epsilon$   \\
\hline
    \end{tabular}
    \caption{Updated Game 2 from EDT Beauty’s perspective.}
    \label{Table_Hitchcock's_Dutch_Book_evidential}
\end{table}

On the contrary, if she is a causal decision theorist, she believes that she can make any choice on the other day. She should therefore align her acceptable betting odds with her credence. In this situation, she is vulnerable to Hitchcock's Dutch Book. Briggs calculates the expectation value of accepting versus not accepting the game each time Beauty wakes up, she obtains \cite{Briggs2010-BRIPAV-5}

$$\mathcal{V}(Accept) - \mathcal{V}(Not ~ Accept) = \epsilon > 0$$

We present a simpler and more intuitive version of Briggs' proof here, with all possible events listed in Table \ref{table_Briggs'_proof_for_CDT}. 

\begin{table}[h]
    \centering
    \begin{tabular}{|c|c|c|c|}
    \hline
\multirow{2}{*}{Heads}& Y &  $10+\epsilon$  \\
% \cline{2-3}
& N & 0   \\
\hline
\multirow{4}{*}{Tails}& YY &  $-20+2\epsilon$  \\
% \cline{2-3}
& YN & $-10+\epsilon$   \\
% \hline
& NY & $-10+\epsilon$   \\
% \hline
& NN & $0$   \\
\hline
    \end{tabular}
    \caption{All possible events corresponding to Beauty's choices in Game 2 of Hitchcock's Dutch Book. The first “Y/N” indicates “Accept/Not accept Game 2 on Monday,” and the second “Y/N” indicates “Accept/Not accept Game 2 on Tuesday.”}
    \label{table_Briggs'_proof_for_CDT}
\end{table}

If the coin lands Head, 

$$\mathcal{V}(Accept|H) - \mathcal{V}(Not ~ Accept|H) = 10 + \epsilon$$

and if it lands Tails,

$$\mathcal{V}(Accept|H) - \mathcal{V}(Not ~ Accept|H) = -10 + \epsilon$$

Suppose Beauty is a Halfer. Then

\begin{align*}
& P(H)(10 + \epsilon) + P(T) (-10 + \epsilon)  \\
& = \frac{1}{2} (10 + \epsilon) + \frac{1}{2} (-10 + \epsilon) \\
& = \epsilon
\end{align*}
 
Thus, if she is a Halfer and follows CDT, she will accept Game 2 every time she wakes up and will be vulnerable to Hitchcock's Dutch Book.

However, we argue that CDT is not a suitable decision theory for the Sleeping Beauty Problem. We raise three major concerns.

(1) Normally, CDT is applicable when some crucial influencing factors of our decision have already occurred in the past and therefore cannot be changed by our current decision. For example, in Newcomb's problem, the amount of money in the boxes has already been determined and cannot be altered by our choice. In SBP, if Beauty is asked on Tuesday, her decision on Monday has already been made, which is a reasonable situation for applying CDT. However, if she is asked on Monday, her decision on Tuesday has not yet been made. This is not a standard situation in which CDT is usually applied. Moreover, both the Halfer and the Thirder positions assign a higher probability to Monday than to Tuesday ($P(Mo)=3/4$ for Halfers and $P(Mo)=2/3$ for Thirders). Given this, it is unclear why the application of CDT should be considered reasonable.

(2) We agree that if Beauty is a Halfer and follows CDT, she will prefer accepting Game 2 to not accepting it. What we disagree with is the claim that this necessarily leads to a sure loss. In Briggs' original proof (see Appendix for more details), although we have $\mathcal{V}(Accept) > \mathcal{V}(Not ~ Accept)$, both $\mathcal{V}(Accept)$ and $\mathcal{V}(Not ~ Accept)$ can be negative when $\epsilon<10/3$ --- This means that we always have $\mathcal{V}(Not ~ Accept)<0$, and we may also have $\mathcal{V}(Accept)<0$ for certain values of $P(Accept*)$, which denotes the probability that Beauty accepts Game 2 on the other day. 

If Beauty is rational, she should not want to face a choice between accepting and not accepting a game when both options lead to a sure loss. In particular, once she has already calculated using CDT that she will accept the game with probability one, it is reasonable to conclude that she would only be willing to enter a situation in which she must choose whether or not to accept Game 2 if $\epsilon > 10/3$. In that case, she would only be willing to bet at odds of 1:2.

One might ask what Beauty can do other than accept or not accept the game. The answer is that she can refuse to participate in the SBP experiment altogether. If the (CDT) Beauty is informed before the experiment that she will be involved in Hitchcock's Dutch Book, she will regard this as a deception by which the experimenter aims to take her money. Once she participates in the SBP experiment, she is guaranteed to lose money regardless of whether she accepts Game 2 or not. If she follows our looser version of CDT in SBP (we call it loose because we allow CDT to be applied even when the influencing factor lies in the future), she should refuse to take part in the SBP experiment, or request that it be stopped immediately when she first hears about Game 2, which would be on Monday. She can justify this by appealing to reasons such as experimental ethics. The memory-erasure procedure makes her suspicion both logical and understandable. In this way, she will not lose any money.

It might also be objected that we cannot use $P(Accept*)=1$ to conclude that $\epsilon$ must be greater than $10/3$ to ensure $\mathcal{V}(Accept)$ to be positive, since $P(Accept*)$ can take any value between 0 and 1. However, this would imply that although Beauty calculates $P(Accept*)=1$ using CDT, she does not need to act following that result. She can therefore still refuse Game 2, which again shows that there is no sure loss in Hitchcock's Dutch Book.

(3) Consider a variant in which Beauty is not woken up twice when the coin lands Tails, but 100 times. In this case, it is more likely that she will make different decisions each time she wakes up. This scenario fits the assumptions of CDT better than those of EDT. But would she then be more willing to follow CDT and accept Game 2 each time she wakes up?

Once she participates in Game 2 more than once, she is guaranteed to incur a loss, and the loss increases the more times she accepts Game 2. Suppose Beauty is not a firm believer in CDT, but merely has a strong preference for it. In this more CDT scenario, her confidence in CDT would be weakened more. This incoherence further suggests that CDT is not the appropriate decision theory to apply to SBP.

In contrast, if Beauty follows EDT and adopts the Halfer position, she can comfortably refuse Game 2 regardless of whether she is woken up twice or 100 times. 

Briggs also presents another Dutch Book showing that EDT is not compatible with the Thirder position \cite{Briggs2010-BRIPAV-5}, as shown in Table \ref{Table_Briggs'_Dutch_Book}.

\begin{table}[h]
    \centering
    \begin{tabular}{|c|c|c|c|}
    \hline
Game & & Heads &  Tails  \\
\hline
1* & Sunday & 15+2$\epsilon$ & -15+$\epsilon$   \\
\hline
2* & Monday/Tuesday & -20+$\epsilon$ & 5+$\epsilon$   \\
\hline
    \end{tabular}
    \caption{Briggs’ Dutch Book demonstrating that EDT is not compatible with the Thirder position \cite{Briggs2010-BRIPAV-5}.}
    \label{Table_Briggs'_Dutch_Book}
\end{table}

If the coin lands Heads, Beauty gains

$$15+ 2 \epsilon -20 + \epsilon=-5 + 3 \epsilon$$

and if the coin lands Tails, she gains

$$-15 + \epsilon + 10 + 2\epsilon =-5 + 3 \epsilon$$

Using EDT, if she accepts the game, she gains

\begin{align*}
&P(H)(-20+\epsilon)+P(T)(10+2\epsilon)\\
&= \frac{1}{3}(-20+\epsilon)+\frac{2}{3}(10+2\epsilon) \\
&= \frac{5}{3} \epsilon
\end{align*}

which is greater than zero, the reward she would receive if she refused Game 2*. Therefore, if Beauty is a Thirder and follows EDT, she will accept the game and suffer a guaranteed loss.

We can now revisit our three concerns in the context of Briggs' Dutch Book. First, EDT can be applied to analyze events that occur both in the past and in the future. Second, both accepting and not accepting Game 2* yield non-negative rewards. Third, if Beauty is woken up more times when the coin lands Tails, the assumptions of EDT become more questionable. We could still use EDT and the new credence $P(H)$ to obtain the result that Beauty will accept the game. This decision is consistent and rational, since $P(H)$ is very small from the Thirder's viewpoint, and the expected value of the games and the gain when the coin lands Tails are both huge. This result seems to support EDT with the Thirder position, but under a more CDT scenario. This also shows that the variant in which Beauty is woken up a large number of times when the coin lands Tails is not analogous to the original SBP and should not be used to analyze it in some sense.

The analysis above shows that EDT is more suitable for solving the SBP than CDT, and therefore $P(H)=1/2$ should be the correct answer.

\section{Conclusion}

We have analysed the analogy between probability in the Many Worlds Interpretation (MWI) and the Sleeping Beauty Problem (SBP), and refuted the viewpoint that uses the dominant position in SBP ($P(H)=1/3$) to question MWI. We have shown that for both the quantum (MWI) SBP and the classical SBP, the correct answer is $P(H)=1/2$.

We show $P(H)=1/2$ for the quantum (MWI) SBP by pointing out an error of renormalization made by the authors of \cite{Hallakoun2013} in Section \ref{subsection_Quantum_SBP}. Then, we prove that this result also holds for four major arguments of the Thirders in the classical SBP: In Section \ref{subsection_Proportion_Argument}, we refute the Proportion Argument by showing that the weight of a possible event cannot be counted as probability. In Section \ref{subsection_Elga's_Variant}, we refute Elga's Variant Argument by extending Burock’s idea \cite{Burock2023} of criticizing this variant for introducing additional information. We clarify this issue using another similar variant and show why the Principle of Indifference cannot be applied, drawing a comparison with d'Alembert's historical mistake. In Section \ref{subsection_Technicolor_Beauty}, we refute Titelbaum's Technicolor Beauty Variant Argument by pointing out the mistake of treating overlapping events as disjoint. Finally, in Section \ref{subsection_Dutch_Book}, we review Briggs' \cite{Briggs2010-BRIPAV-5} analysis of the Halfer and Thirder positions in relation to evidential decision theory (EDT) and causal decision theory (CDT), and show that CDT is not suitable for dealing with SBP. For all four arguments, we conclude that $P(H)=1/2$ is the only correct answer. Although some of our analyses are based on previous literature, we have developed new persuasive arguments in each part. All of our proofs use rigorous mathematical calculations, avoiding the ambiguous philosophical debates that are very common in discussions of SBP.

Our results not only show that MWI is coherent with classical probability theory, especially in situations closely analogous to it, namely cases with $\tilde{P}(A)=1$ but $P(A)<1$ for some event $A$, but also challenge the dominant position in the classical Sleeping Beauty Problem. Even if readers are not fully convinced by our conclusion that the Halfer position is the correct one, we have at least shown that the problem cannot be regarded as closed and that the Thirder position cannot be taken as the final answer. 

We also hope to stimulate interest in analysing SBP with other interpretations of quantum mechanics, which could provide a promising way of testing those interpretations, as in the case of MWI.

\section{Acknowledgements}

We thank Vlatko Vedral for valuable advice on this work, and Lev Vaidman for helpful discussions related to his work on this problem. We also thank Aditya Iyer and Cosmin Andrei for inspiring discussions.

\newpage
\onecolumngrid

\bibliography{ref.bib}

\newpage
\onecolumngrid

\section*{Appendix}

\subsection{Briggs' Proof for Beauty Accepting Game 2 as a Halfer and Follows CDT}

Briggs calculates the expectation value of Beauty (assuming she is a Halfer and a causal decision theorist) either accepting or not accepting Game 2 \cite{Briggs2010-BRIPAV-5}:

\begin{align*} 
\mathcal{V}(Accept) 
&= P(H)\cdot (10 + \epsilon) \\
&\quad + P(T)\cdot \bigl[ P(Accept*) \cdot (-20 + 2\epsilon) \\
&~~~~~~~~~~~~~~~~~ + P(Not ~ Accept*) \cdot (-10 + \epsilon) \bigr] \\
&= 5 + \frac{\epsilon}{2}+\frac{1}{2}[-10+\epsilon+P(Accept*) \cdot (-10+\epsilon)] \\
&= \epsilon + P(Accept*) \cdot (-5 + \frac{\epsilon}{2}) 
\end{align*}

\begin{align*} 
\mathcal{V}(Not ~ Accept) 
&= P(H)\cdot 0 \\
&\quad + P(T)\cdot \bigl[ P(Accept*) \cdot (-10+\epsilon) \\
&~~~~~~~~~~~~~~~~~ + P(Not ~ Accept*) \cdot 0 \bigr] \\
&= P(Accept*) \cdot (-5 + \frac{\epsilon}{2})
\end{align*}

Since $\mathcal{V}(Accept) - \mathcal{V}(Not ~ Accept) = \epsilon > 0$, Beauty will accept the game every time she wakes up.

Additionally, we always have $\mathcal{V}(Not ~ Accept) \le 0$ if $\epsilon < 10$, and it is also possible that $\mathcal{V}(Accept) < 0$ for some values of $P(Accept*)$ when $\epsilon < 10/3$. This makes the game close to a deception and may cause problems for Hitchcock's Dutch Book (see Section \ref{subsection_Dutch_Book}).

\end{document}